# Free Space Optics for Next-Generation Satellite Networks

Aizaz U. Chaudhry and Halim Yanikomeroglu
Department of Systems and Computer Engineering, Carleton University, Ottawa, Canada – K1S 5B6

**Abstract–Free space optics (FSO) refers to optical wireless communications in outdoor environments. The aim of this paper is to analyze the role that FSO is envisaged to play in the creation of next-generation satellite networks. To begin with, the reader is introduced to the types of FSO links and functionalities of a basic FSO system. Next, a comparison of FSO and radio frequency (RF) technologies for inter-satellite links (ISLs) is provided, including a comparison between FSO and RF links when employed between low Earth orbit satellites. After that, the types of FSO or laser ISLs are considered, and the challenges in establishing laser ISLs, the properties of laser ISLs, and the capabilities of terminals for laser ISLs are discussed. Then, the parameters of a satellite constellation are highlighted, and different aspects of SpaceX's upcoming mega-constellation Starlink are explored. In addition, the optical wireless satellite network that is created by utilizing laser ISLs is examined. Finally, a use case is investigated for next-generation optical wireless satellite networks that are envisioned by the mid to late 2020s.**

WHAT IS FREE SPACE OPTICS?

In optical wireless communications (OWC), an optical carrier is employed to send information over an unguided channel. In optical fiber communications and OWC, lasers beams are used for data transfer. Unlike optical communications over fiber, however, data in OWC is sent over an unguided channel (or medium), such as the atmosphere or the vacuum in space, instead of a guided medium like optical fiber.

There are two main types of OWC: indoor and outdoor. Outdoor OWC is referred to as free space optics (FSO) or free space optical communications. FSO can be further divided into four link sub-types: terrestrial, non-terrestrial (or aerial), space, and deep-space. Examples of terrestrial FSO are building-to-building FSO links. Non-terrestrial FSO links include ground-to-unmanned aerial vehicles (UAVs), ground-to-high altitude platform systems (HAPSs), and HAPS-to-HAPS. Ground-to-satellite, satellite-to-ground, and satellite-to-satellite FSO links are instances of space FSO. Deep-space FSO can be an FSO link between the Earth and a spacecraft in deep-space like Galileo. Satellite-to-satellite FSO links, which are also referred to as laser inter-satellite links (ISLs), are the focus of this paper.

The first significant breakthrough that laid the foundation for today's FSO came with the discovery of the first working laser in 1960. The first optical or laser ISL was demonstrated in 2008 between two low Earth orbit (LEO) satellites, Terra SAR-X and NFIRE. The capacity of this laser ISL was 5.5 Gbps, and it was established at an inter-satellite distance of 5,500 km at a speed of 25,000 km/hr. A brief history of the developments in FSO, mainly with respect to space, is provided in [1].

In an FSO link, an optical signal is transmitted from an optical transmitter to an optical receiver over the atmosphere or the vacuum in space. Unobstructed line of sight is required between a transmitter and receiver for FSO communications. A basic FSO system is described in [2]. At the transmitter, the input data is converted into a modulated electrical signal that controls the intensity of light coming out of the laser by controlling the laser current in the driver. The telescope at the transmitter focuses this modulated laser beam at the receiver's telescope. This optical signal travels through the medium and is received at the receiver's telescope. The photodetector at the receiver converts this optical signal to an electrical signal, which is then demodulated to retrieve the transmitted data. An optical filter at the receiver eliminates the background solar radiation.

This paper aims to provide an analysis of the vital role that FSO is envisioned to play in next-generation satellite networks. After introducing the types of FSO links and functionalities of a basic FSO system, a comparison of FSO and RF technologies for ISLs is provided, including a comparison between FSO and RF links for LEO satellites. Next, the types and properties of FSO or laser ISLs are considered, and the challenges in establishing laser ISLs and the capabilities of terminals for laser ISLs are explored. After that, the basic parameters of a satellite constellation and various aspects of SpaceX's upcoming mega-constellation Starlink are highlighted. Then, the optical wireless satellite network formed from employing laser ISLs is examined. Lastly, a use case for next-generation optical wireless satellite networks as providers of low-latency communications over long distances is investigated.

Although the mentioned aspects have been analyzed in the literature individually [1–7], to the best of our knowledge, this is the first paper to analyze all these aspects of the next-generation optical wireless satellite networks together at one place within a framework. Furthermore, a use case analyzing the suitability of next-generation optical wireless satellite networks for low-latency communications does not exist in the literature.

COMPARISON OF FSO AND RADIO FREQUENCY FOR INTER-SATELLITE LINKS

Both FSO and radio frequency (RF) signals are instances of electromagnetic waves. RF is a well-established wireless communications technology that is currently in its fifth generation for mobile communications [8]. However, FSO is emerging as a viable alternative to RF [9]. For point-to-point wireless communication in space, especially satellite-to-satellite links or ISLs, the benefits of an FSO link compared to an RF link heavily outweigh the drawbacks, and a comparison of the two is presented in Table 1 [3]. Typically, the beam spread or beam divergence of an RF signal is 1,000 times greater than that of an FSO signal [10].

Compared to RF links, higher frequency and higher bandwidth of FSO links as well as the characteristics of their laser beam present the following significant advantages: higher data rates; smaller antenna sizes leading to lesser weight and lesser volume; narrower beams resulting in no interference and higher security; and lower transmit power due to lower beam spread and higher directivity. Due to their smaller size, weight, volume, and power requirement, FSO terminals require less onboard satellite resources and can be easily integrated into satellite platforms. The smaller form factor of FSO terminals also helps in reducing satellite launching and deployment costs.

Laser ISLs are envisioned between satellites in upcoming LEO satellite constellations. A comparison between FSO and RF links when employed for an LEO-to-LEO satellite link is provided in [4]. The RF links operate in Ka and mm-wave bands, and the transmit power is considered as

TABLE 1. FSO VS. RF FOR INTER-SATELLITE LINKS

| Characteristic | FSO Link | RF Link |
|---|---|---|
| Spectrum | near infrared and visible light band | Ka, Ku, and mm-wave band |
| Bandwidth | in THz | in GHz |
| Wavelength | in nm | thousands of times larger than the optical wavelength |
| Antenna size | smaller antenna size due to smaller wavelength | much bigger antenna size due to much larger wavelength |
| Beam spread | much narrower | typically 1,000 times more |
| Directivity | very high due to much smaller wavelength | very low due to much larger wavelength |
| Power consumption | more received power for a given transmitted power | less received power for a given transmitted power |
| Spectrum allocation | unlicensed | licensed |
| Security | very difficult to intercept due to high directivity | easy to intercept |
| Interference | no interference due to narrow beam width | severely degrades performance |
| Solar radiation | lead to poor performance | do not affect performance |
| Line-of-sight obstructions | severely impact performance | severely impact performance at mm-wave band |

10 W, 20 W, and 50 W for FSO, mm-wave and Ka links, respectively. At a data rate of 2.5 Gbps over an inter-satellite distance of 5,000 km, the RF ISL in either Ka or mm-wave bands requires at least 19 times the antenna diameter, and more than twice the onboard power and mass compared to an FSO ISL.

## TYPES, CHALLENGES, PROPERTIES, AND TERMINALS FOR LASER ISLs

Depending on the location of the satellites, FSO or laser ISLs can be classified into three types: intra-orbital plane ISL, which is between two satellites in the same orbital plane; inter-orbital plane ISL, which is between satellites in two different orbital planes; and inter-orbit ISL, which is between satellites in two different orbits. Within an orbital plane, satellites at the same altitude move in the same plane and in the same direction. In different orbital planes, satellites are at the same altitude but they move in different planes in similar or crossing paths. Satellites in different orbits are at different altitudes and may or may not move in the same direction. These different types of laser ISLs are illustrated in Fig. 1. The solid red/blue lines indicate intra-orbital plane ISLs, the solid yellow line indicates an inter-orbital plane ISL, and the solid green line indicates an inter-orbit ISL.

Satellites in an orbital plane move with the same velocity (i.e., same speed and direction), and intra-orbital plane ISLs are relatively easy to establish and maintain. Inter-orbital plane ISLs with satellites in adjacent orbital planes are more difficult to establish. Different orbital planes have the same altitude, and satellites in these orbital planes move at the same speed. However, the direction of satellites in adjacent orbital planes is slightly different, which leads to different relative velocities of satellites. Also, satellites in crossing orbital planes or satellites in different orbits move at high relative velocities and suffer from challenges like Doppler shift, point-ahead angle (PAA), and acquisition, tracking, and pointing (ATP). Intra-orbital plane ISLs and inter-orbital plane ISLs with satellites in adjacent orbital planes are generally stable. Inter-orbital plane ISLs with satellites in crossing orbital planes are make-break or intermittent in nature and cannot last for long durations.

The small beam width is the main advantage of FSO as it eliminates interference from neighbors. However, in laser ISLs, the small beam width turns into a disadvantage. Pointing a laser beam at a moving satellite from another moving satellite becomes a major challenge due to the narrow beam divergence of the laser beam and the different relative velocities of the satellites. Hence arises the demand for a very precise ATP system on board a satellite platform for the laser beam originating from one satellite to effectively connect to another satellite. Suppose satellite *A* receives a laser beacon from satellite *B*. Due to their different relative velocities, *A*'s response signal to *B* has to be offset from *B*'s original location so that it is received by *B* at the appropriate spatial and temporal location. This pointing offset is referred to as point-ahead angle and is taken care of by the ATP system. The different relative velocity of the sender and the receiver result in a change in the frequency of the received signal, which is referred to as Doppler shift. This shift is present for a pair of satellites in adjacent orbital planes and is higher for two satellites in crossing orbital planes or in different

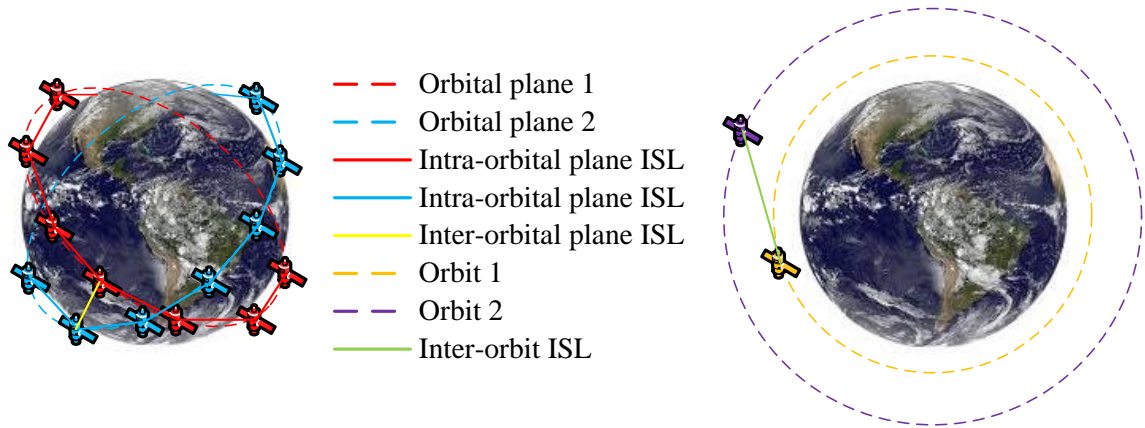

Fig. 1. Types of laser ISLs.

orbits. This has to be addressed to establish a reliable laser ISL [5].

The high capacity and low latency requirements of next-generation satellite networks necessitate laser ISLs. In the absence of such ISLs, a long-distance connection between two cities, such as an inter-continental connection between New York and London, must ping-pong between ground stations and satellites. The main properties of laser ISLs can be summarized as follows: setup times for laser ISLs vary from a few seconds to tens of seconds. These setup times and the intermittent nature of laser ISLs with satellites in crossing orbital planes or with satellites in different orbits make such ISLs undesirable. Laser ISLs are mainly constrained by visibility. For example, the maximum ISL range for an LEO satellite operating at an altitude of 550 km is 5,014 km. Laser ISLs offer capacities in Gbps and interference is not an issue for these ISLs [6].

Laser terminals for establishing laser ISLs will be absolutely critical in the formulation of a global communications network in space by inter-connecting hundreds of satellites in next-generation satellite networks, such as the upcoming LEO satellite constellations. Two such terminals have already been developed by Tesat [11] while others are currently under development by companies like Mynaric [12] and General Atomics [13]. A summary of the capabilities of these terminals in terms of their specifications is given in Table 2.

SATELLITE CONSTELLATION BASICS AND AN UPCOMING EXAMPLE

A uniform satellite constellation, where satellites have circular orbits, can be defined by four parameters: inclination, which is the angle between the orbit of the satellite constellation and the Equator; altitude, which is the height of the orbit from the surface of the Earth and is measured from the sea level; number of orbital planes; and number of satellites per orbital plane. In a circular orbit, the apogee and the perigee of the satellite (i.e., the highest altitude and the lowest altitude of the satellite relative to Earth's surface, respectively) are the same. In a uniform satellite constellation, the spacing between orbital planes is the same and so is the spacing of satellites within an orbital plane.

There can be several orbital planes within an LEO (or very low Earth orbit (VLEO)) satellite constellation, and each orbital plane can have several LEO (or VLEO) satellites. Different LEO and/or VLEO satellite constellations at different altitudes collectively constitute a mega-constellation. Let us consider the example of an LEO satellite constellation with the following parameter set: {53º, 550 km, 40, 40}. This indicates an orbital inclination of 53º, an orbital altitude of 550 km, a total of 40 orbital planes, and a total of 40 satellites within each orbital plane. Assuming this constellation to be uniform, the spacing between the orbital planes will be 9º, and the spacing between satellites in each orbital plane will also be 9º. All orbital planes in the constellation have the same inclination with reference to the Equator.

Starlink is SpaceX's upcoming mega-constellation that is expected to be comprised of approximately 12,000 satellites in different orbits or constellations in LEO and VLEO orbits. The plan for the deployment of LEO constellations is provided in Table 3 [14]. This will consist of a total of 4,425 satellites in five LEO constellations. It will enable SpaceX to provide full and continuous global coverage for a wide range of broadband and other communications services for users worldwide. These satellites will be equipped with phased array beam-forming antennas to make efficient use of Ku and Ka band spectrum resources.

In the modified phase I for constructing Starlink shown in Table 4 [15], SpaceX is currently in the process of deploying an LEO constellation of 1,584 satellites with a parameter set of {53º, 550 km, 24, 66}. In the original authorization obtained by SpaceX from FCC, phase I consisted of 1,600 satellites with a parameter set of {53º, 1,150 km, 32, 50} [15]. This new or modified phase I constellation is illustrated in Fig. 2(a). Assuming this constellation to be uniform, the spacing between the orbital planes is 15º, and the spacing between satellites in each orbital plane is 5.45º in this figure. The LEO satellites that have currently been deployed by SpaceX as part of phase I are not equipped with laser terminals and cannot establish intra-orbital plane or inter-orbital plane laser ISLs. SpaceX is looking to incorporate this capability in its Starlink satellites beginning in late 2020.

TABLE 2. LASER TERMINALS FOR LASER ISLs

| Company | Terminal | Capacity | Standard Compatibility | Link Distance | ATP Capability |
|---|---|---|---|---|---|
| Mynaric | CONDOR | 10 Gbps | Ethernet IEEE 802.3 | 8,000 km | Available |
| Tesat | LCT 135 | 1.8 Gbps | - | 80,000 km | - |
| Tesat | SmartLCT | 1.8 Gbps | - | 45,000 km | - |
| General Atomics | 1,550 nm LCT | 5 Gbps | - | 2,500 km | - |

In phase I, SpaceX is building and deploying their first-generation (1G) satellites with a simplified design. For example, these 1G satellites will use Ku band spectrum for links between satellites. New features, like laser ISLs, will be added to subsequent generations of satellites. Similarly, their 1G satellites will use parabolic antennas to link to ground stations, and phased array solutions will be introduced in later generations [15].

SpaceX plans to build hundreds of ground stations (or gateway sites) around the world so that its Starlink satellites have a visible ground station to communicate with from anywhere in their orbit. The number of gateways may increase with user demand. During initial deployment, SpaceX is expecting to deploy approximately 200 gateways in the United States. SpaceX is planning to install a few telemetry, tracking, and control (TT&C) Earth stations all around the world. Initially, two such stations will be installed in the United States, including one on the East coast and one on the West coast [15].

The useful lifetime of a Starlink satellite is expected to be five to seven years. When satellites reach their end of life, SpaceX plans to dispose of them through atmospheric reentry. Operating at lower altitudes can be advantageous especially in unplanned scenarios. In a worst case scenario, when a satellite fails and has no attitude control, it will take around five years for a satellite at 550 km to enter Earth's atmosphere under natural orbital decay compared to hundreds of years at 1,150 km. The reliability of satellites also increases at lower altitudes due to the reduction in the intensity of radiation, and there is a lower risk of failure induced by radiation. Less propellant is required to place the satellite in orbit at a lower altitude, which leaves more fuel in the satellite for keeping it in its desired orbit [15].

OPTICAL WIRELESS SATELLITE NETWORKS

In its FCC filings, SpaceX mentioned five 1.5 kg silicon carbide communication components while discussing space debris issues when its Starlink satellites reach end of life and need to be de-orbited [14]. Silicon carbide is used in mirrors in laser terminals for laser ISLs. This indicated that SpaceX would equip its Starlink satellites with five laser ISLs so that they could connect to other satellites within a constellation to form an optical wireless satellite network.

Four of these laser ISLs will likely be used to connect to nearby satellites in the same and adjacent orbital planes. The obvious candidates for these ISLs are the two neighbors of a satellite in the same orbital plane, one in the front and one at the back, and the two neighbors in two adjacent orbital planes, one at the right and one at the left. At any one region in space above the Earth, half of the satellites within the constellation are moving in a northeasterly direction while the other half are moving in a southeasterly direction. The fifth laser ISL is likely intended to connect to a satellite in a crossing orbital plane [7]. The satellites in the Starlink constellation at 550 km altitude travel at speeds of approximately 7.6 km/s. It is challenging for the laser ISL of a northeast-bound satellite to track and connect to a southeast-bound satellite and then rapidly switch to a new crossing satellite as the old one moves away [7]. In a later FCC filing, SpaceX revised the number of ISLs per satellite to four [6]. This could be due to the difficulty in engineering this fifth laser ISL.

TABLE 3. SPACEX'S PLAN FOR LEO CONSTELLATIONS

| Parameter | Phase I Deployment | Phase II Deployment | | | |
|---|---|---|---|---|---|
| Inclination | 53º | 53.8º | 74º | 81º | 70º |
| Altitude | 1,150 km | 1,110 km | 1,130 km | 1,275 km | 1,325 km |
| No. of Orbital Planes | 32 | 32 | 8 | 5 | 6 |
| No. of Satellites per Orbital Plane | 50 | 50 | 50 | 75 | 75 |
| Total Satellites | 1,600 | 2,825 | | | |

TABLE 4. SUMMARY OF PROPOSED MODIFICATION FOR PHASE I

| Parameter | Original Authorization | Proposed Modification |
|---|---|---|
| No. of Orbital Planes | 32 | 24 |
| No. of Satellites per Orbital Plane | 50 | 66 |
| Total Satellites | 1,600 | 1,584 |
| Altitude | 1,150 km | 550 km |
| Inclination | 53º | 53º |

Connecting with front and back neighbors within orbital planes provides southwest ↔ northeast and northwest ↔ southeast routing paths within the two groups of satellites moving in northeasterly and southeasterly directions, respectively. Instead of connecting with immediate left or right neighbors in adjacent orbital planes, a good idea could be to connect to adjacent orbital plane neighbors in north-south or east-west directions. Providing east-west connectivity may be preferable to north-south connectivity in phase I of Starlink as it would benefit most of the population in developed countries including the United Sates [7].

The optical wireless satellite network that is created by utilizing each satellite's four laser ISLs in this manner provides a good mesh network, as illustrated in Fig. 2(b). However, there are two distinct meshes that exist within the network, the red colored one between the group of satellites moving northeast and the blue colored one between the group moving southeast. There is no local connectivity between these meshes. Traffic can be routed without switching between the two meshes. Using an additional or fifth laser ISL to create inter-mesh links or links between crossing orbital planes can improve routing options. However, such links are expected to be down frequently as they are re-established from one crossing satellite to another [7].

NEXT-GENERATION SATELLITE NETWORKS – A USE CASE

The refractive index (or index of refraction) of a medium is the ratio of the speed of light in vacuum to the speed of light in that medium [16]. The higher the refractive index, the slower the light travels through that medium. If a medium has a refractive index of 2, the light will travel through it at 1/2 the speed of light in vacuum. The speed of light in vacuum is ~ $3 \times 10^8$ m/s and is generally denoted by $c$ [17].

Typical optical fibers are made of glass with an index of refraction of ~ 1.5. Therefore, the speed of light in optical fiber is ~ 2/3 of the speed of light in vacuum or ~ $c/1.5$. This means that the speed of light in vacuum is ~ 50% higher than in optical fiber. This interesting and well-known fact gains critical importance in optical wireless satellite networks. It gives them an advantage over terrestrial optical fiber networks in terms of latency or propagation delay when the data communications takes place

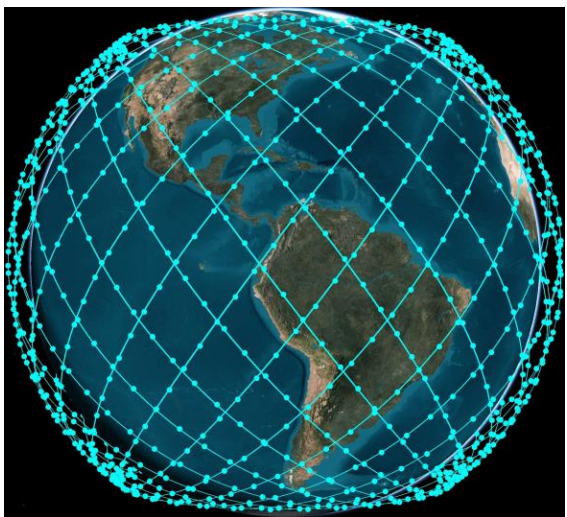 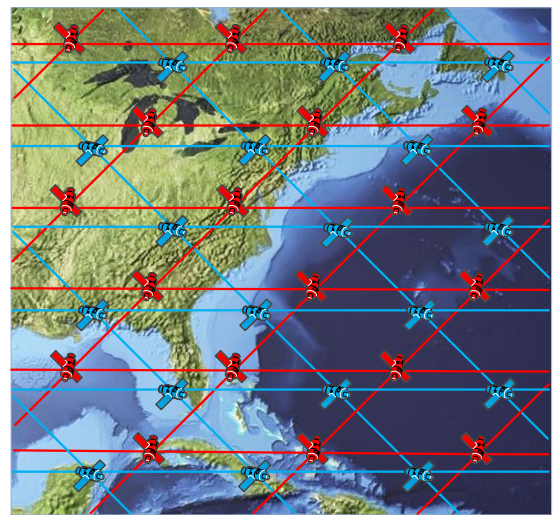

(a)          (b)

Fig. 2. (a) SpaceX's LEO constellation for its new phase I; (b) Optical wireless satellite network created using four laser ISLs per satellite.

over long distances of more than 3,000 km [7].

A next-generation satellite network that may come into existence by the mid to late 2020s will be based on such an optical wireless satellite network that is created by using laser ISLs in satellite constellations like phase I of Starlink. Providing low-latency communications over long distances may turn out be the primary use case of this next-generation satellite network. The cost of creating and maintaining such a network can easily be recovered by providing low latency communications as a premium service to the financial hubs around the world. As per Information Week Magazine, it is estimated that in trading stocks at the stock exchange, a 1 millisecond advantage can be worth $100 million a year to a single major brokerage firm [18]. It is not hard to imagine that an advantage of a few milliseconds in financial stock markets may result in billions of dollars of revenues for these financial firms. These firms are already looking for technological solutions such as high speed communications networks to reduce latency, and a low latency next-generation satellite network may provide the perfect solution.

In phase I, the LEO satellites of Starlink's upcoming constellation will orbit the Earth at an altitude of 550 km. To communicate over such a next-generation satellite network, a combined ingress and egress distance of 1,100 km will be involved from Earth to ingress satellite (entry point of the satellite network) and from egress satellite (exit point of the satellite network) back to Earth. This extra distance to use the satellite network prevents lower latencies compared to the terrestrial optical fiber network over shorter distances. However, the extra latency resulting from this extra distance can easily be offset by communicating over longer distances between cities (or financial centers) around the world at the speed of light over next-generation satellite networks.

A comparison between next-generation satellite networks and existing terrestrial optical fiber networks in terms of latency, or more specifically propagation delay, is illustrated in Table 5. The scenario illustrating this comparison is shown in Fig. 3. The refractive index of a Corning® single-mode optical fiber suitable for communications over long distances is 1.4675 at 1,310 nm [19] and the speed of light in this optical fiber is $c/1.4675$. The exact value of the speed of light in vacuum or $c$ is 299,792,458 m/s [20]. In this example, we consider the exact value of the speed of light in vacuum for a next-generation satellite network. We consider the speed of light in optical fiber in a terrestrial network to be $c/1.4675$, i.e., 204,287,876 m/s. The parameters of SpaceX's phase I of its Starlink constellation are considered for the satellite network in this scenario, i.e., an altitude of 550 km and a uniform spacing of 5.45° between the 66 satellites within an orbital plane. The radius of Earth is 6,378 km.

The linear distance between the first two LEO satellites $S$ and $T$ within the orbital plane is calculated using the formula for the length of a chord, i.e., $2\ r\ \sin((\theta/2)\ (\pi/180))$, where $\theta$ (the angular spacing between $S$ and $T$ in degrees) is 5.45°, and $r$ is 6,378 km + 550 km. This is the hop distance between $S$ and $T$ or between any two consecutive satellites in the orbital plane for that matter. This distance is multiplied by two to find the distance between $S$ and its second hop, by three to find the distance to its third hop, and so on. To calculate the end-to-end distance for *S* and its $n^{th}$ hop, 1,100 km is added to their hop distance to account for the distance between Earth and *S* and the distance between the $n^{th}$ hop of *S* and Earth.

TABLE 5. COMPARISON OF LATENCY

| No. of Hops | $\theta$ (degrees) | Next-Generation Satellite Network | | | Terrestrial Optical Fiber Network | | |
|---|---|---|---|---|---|---|---|
| | | Hop Distance (km) | End-to-End | | Hop Distance (km) | End-to-End | |
| | | | Distance (km) | Latency (ms) | | Distance (km) | Latency (ms) |
| 1 | 5.45° | 659 | 1,759 | 5.87 | 607 | 607 | 2.97 |
| 2 | 10.90° | 1,317 | 2,417 | 8.06 | 1,213 | 1,213 | 5.94 |
| 3 | 16.35° | 1,976 | 3,076 | 10.26 | 1,820 | 1,820 | 8.91 |
| 4 | 21.80° | 2,635 | 3,735 | 12.46 | 2,427 | 2,427 | 11.88 |
| 5 | 27.25° | 3,294 | 4,394 | 14.66 | 3,033 | 3,033 | 14.85 |
| 6 | 32.70° | 3,952 | 5,052 | 16.85 | 3,640 | 3,640 | 17.82 |
| 7 | 38.15° | 4,611 | 5,711 | 19.05 | 4,247 | 4,247 | 20.79 |
| 8 | 43.60° | 5,270 | 6,370 | 21.25 | 4,853 | 4,853 | 23.76 |
| 9 | 49.05° | 5,929 | 7,029 | 23.45 | 5,460 | 5,460 | 26.73 |
| 10 | 54.50° | 6,587 | 7,687 | 25.64 | 6,067 | 6,067 | 29.70 |

The distance between the first terrestrial optical fiber relay station *P* and its next hop *Q* along the surface of the Earth is calculated using the formula for the length of an arc, i.e., $2 \pi R (\theta/360)$, where $\theta$ is 5.45º and *R* is the radius of the Earth. To find the distance between *P* and its second hop, third hop, and so on, $\theta$ is 10.90º, 16.35º, and so on. The hop distance and the end-to-end distance for the terrestrial network are the same. To calculate latency for satellite and terrestrial networks, the end-to-end distance is divided by the speed of light in vacuum and the speed of light in optical fiber, respectively. The satellite network operating at an altitude of 550 km begins to outperform the terrestrial optical fiber network in terms of latency or propagation delay starting at five hops when the terrestrial hop distance exceeds 3,000 km.

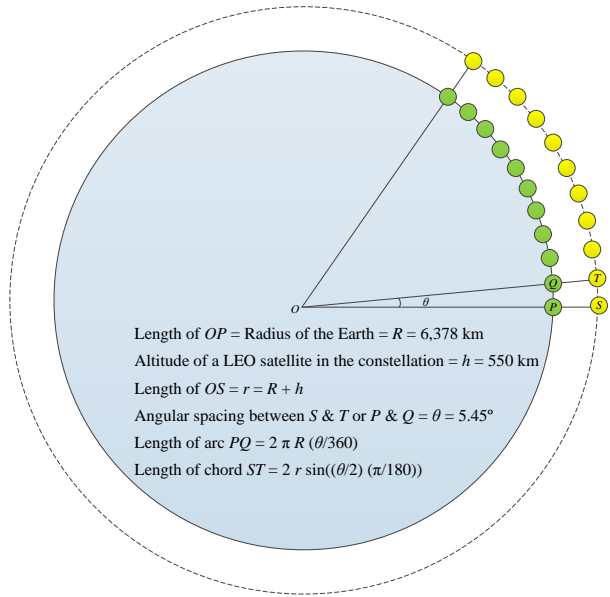

Fig. 3. A scenario for illustrating the latency comparison between a next-generation satellite network and a terrestrial optical fiber network. Earth is shown in blue, LEO satellites in yellow, and terrestrial optical fiber relay stations in green.

The four components of the end-to-end delay in the network include transmission delay, processing delay, queuing delay, and propagation delay [21]. In an optical communication network, such as an optical wireless satellite network or a terrestrial optical fiber network, the propagation delay is the delay introduced by the optical transmission of the signal along the fiber or vacuum. It is directly proportional to the end-to-end distance and becomes very significant for communications over long distances [22]. For the comparison in Table 5, we define latency (or end-to-end latency) as the propagation delay that is measured in only one direction, i.e., from the source to the destination.

## SUMMARY

FSO has experienced an immense technological evolution over the last six decades. It is emerging as a viable alternative to RF for inter-satellite links. Compared to RF terminals, FSO or laser terminals require less onboard satellite resources due to their smaller form factor, and they are easy to integrate into satellite platforms. Laser terminals for laser ISLs are still in their infancy and are expected to offer capacities of up to 10 Gbps. Laser ISLs offering capacities in the hundreds of Gbps will need to be developed to establish a true global communications network in space.

Laser ISLs with satellites in crossing orbital planes and with satellites in different orbits are currently considered undesirable. However, they will be crucial for effective networking within a satellite constellation and between different constellations to achieve mega-constellation satellite networks. Current setup times for laser ISLs are prohibitive and need to be significantly reduced through technological advancement. Efficient ATP systems need to be developed to support reliable ISLs among crossing orbital planes and among different orbits.

Four laser ISLs are expected per Starlink satellite, and SpaceX is looking to equip its satellites with this capability by late 2020. This number of laser ISLs restricts the connectivity to two satellites in the same orbital plane and to two satellites in adjacent orbital planes. Connecting to adjacent orbital plane neighbors in east-west directions is expected as this would benefit the population in the developed world, especially in the United States. Six ISLs per satellite are needed to provide north-south connectivity in addition to east-west connectivity. A next-generation satellite will need to be equipped with several laser ISLs to provide robust connectivity within the network and this will be vital for ensuring low latency paths within optical wireless satellite networks.

The higher speed of light in vacuum gives optical wireless satellite networks operating at 550 km altitude a critical advantage over terrestrial optical fiber networks in terms of latency when communication distances are greater than 3,000 km. The creation of fully functional next-generation optical wireless satellite networks is expected by the mid to late 2020s, and providing low-latency communications over long distances may turn out to be their primary use case.

ACKNOWLEDGEMENTS

This work has been supported by the National Research Council Canada's (NRC) High Throughput Secure Networks program (CSTIP Grant #CH-HTSN-608) within the Optical Satellite Communications Canada (OSC) framework. The authors would like to thank AGI for the Systems Tool Kit (STK) platform.

ABOUT THE AUTHORS

**Aizaz U. Chaudhry** is a Research Associate in the Department of Systems and Computer Engineering at Carleton University. He is interested in research on the application of machine learning and optimization in wireless networks. He is a Senior Member of the IEEE. Contact him at auhchaud@sce.carleton.ca.

**Halim Yanikomeroglu** is a Full Professor in the Department of Systems and Computer Engineering at Carleton University. His research interests cover many aspects of wireless technologies with special emphasis on wireless networks. He is a Fellow of the IEEE. Contact him at halim@sce.carleton.ca.